\documentclass[10pt, english,pra,aps,floatfix,superscriptaddress,
showpacs, notitlepage]{revtex4-1}
\usepackage[T1]{fontenc}
\usepackage{color}
\usepackage{hyperref}
\usepackage{datetime}
\usepackage{babel}
\usepackage{verbatim}
\usepackage{refstyle}
\usepackage{xcolor}
\usepackage{amsthm}
\usepackage{amsmath}
\usepackage{amsfonts}
\usepackage{graphicx}
\usepackage{color}
\usepackage[braket, qm]{qcircuit}
\usepackage{slashbox}

\def\be{\begin{equation}}
\def\ee{\end{equation}}
\def\ba{\begin{eqnarray}}
\def\ea{\end{eqnarray}}
\def\la{\langle}
\def\ra{\rangle}

\def\h{\hskip 1cm}

\def\lo{\longrightarrow}

\begin{document}

\makeatletter

\title{Entanglement-assisted communication in the absence of shared reference frame}

\author{Ali Beheshti}
\email{alibeheshti.edu@gmail.com}
\affiliation{Department of Physics, Sharif University of Technology, Tehran, Iran}
\author{Sadegh Raeisi}
\email{sadegh.raeisi@gmail.com}
\affiliation{Department of Physics, Sharif University of Technology, Tehran, Iran}
\author{Vahid Karimipour}
\email{vahid@sharif.edu}
\affiliation{Department of Physics, Sharif University of Technology, Tehran, Iran}
\date{\today}
\begin{abstract}
Alice wants to convey the value of a 
parameter to Bob with whom she does not share a reference frame. What physical object can she use for this task?
Shall she encode this value into the angle between two physical vectors such as the angle between two spins?
Can she benefit from using entanglement? 
We investigate  these questions here and show that an entangled state of two qubits 
has three parameters that are invariant under changes of the reference frame. 
We also calculate
 for specific examples,  
the average information gain for different circumstances,  
where  one of these parameters 
is used for communication. 
We compare our result with the special case of separable states and find that entanglement enhances the information gain.
\end{abstract}
\maketitle

\section{Introduction}
An essential assumption behind many quantum protocols is the existence of a reference frame. For instance, whenever the spin angular momentum of a spin 1/2 particle is to be measured, the result 
is interpreted based on a reference for spatial directions and so the quantum state assigned to the particle depends on the reference frame. Moreover, many quantum information processing tasks rely on the fact that spatially separated parties have access to a shared frame of reference \cite{proto1,proto2,proto3,proto4}. For instance in the quantum teleportation protocol, Alice performs a local measurement and sends the outcome of the local measurement via a classical message to Bob and then, Bob would perform a local unitary operation based on Alice's measurement outcomes. This will provide him with the state planned to be teleported. For the protocol to work, Alice and Bob need to share a spatial frame of reference, otherwise Bob would perform a wrong operation and teleportation would fail. \\

The information about a direction in space or a moment in time is recognized as unspeakable information which means that it is not indifferent to the physical nature of the carrier and certain material objects must be used to convey this information \cite{unspeak}. In other words, it is called unspeakable since a reference frame is required, to which this information is defined and cannot be clearly presented by a string of classical bits. One might not always possess a physical system that is capable of carrying this kind of information or can act as a reference frame. 
Sometimes it is even impossible to find the perfect description of one's local reference frame,  
either due to misalignment or lack of precision 
or adequate stability.
Thus, a shared reference frame (SRF) is often regarded as a resource \cite{rfrev}. 
The restriction of lacking shared reference frames stimulated an interesting topic of research that attempts to develop a framework for investigating the manipulation of systems that can serve as a reference frame and quantifying the resource that they can provide. This framework is called the resource theory of quantum reference frames \cite{rta1} and is recently treated more generally in the quantum resource theory of asymmetry \cite{rta2, rta3, rta4, rta5, rta6}. In these theories, any quantum system that is aligned with some reference frame is a resource to others with different frames of reference. These resources are useful as a substitute for a classical reference frame and 
 play the same role that entangled states do under the restriction of local operations and classical communications in the resource theory of entanglement \cite{rte1,rte2,rte3}. 
 \\

As mentioned in the previous paragraph aligning reference frames is quite an intricate matter. Several researches have been devoted to exploring techniques that can efficiently establish a SRF either by sending information about directions 
\cite{dirshar1,dirshar2,dirshar3,dirshar4,dirshar5} or by finding the unitary operation that relates the reference frames of two parties \cite{finduni1,finduni2}, or even by employing shared entangled states to interestingly substitute a SRF in certain communication tasks \cite{rezaz2} or securely establish a SRF between parties \cite{rezaz1}. \\

The other approach to obviate the problem of lacking a SRF, that is the focus of this paper, is to circumvent the difficulty by relational encoding. It is shown that under the absence of SRF between parties, one can suppose that their descriptions and operations undergo a random unitary channel \cite{rfrev}. To illustrate, if Alice describes her system by the state $\rho$, the state to someone who has no SRF with Alice, prior to any measurements, is described by
\be
\tilde{\rho}={\cal{E}}[\rho]=\int_G D(g) \rho D^\dag(g) dg, 
\ee
where $dg$ is the Haar measure over the group of transformations $G$ relating the frames of reference to each other and $D(g)$ is the representation of the group on the Hilbert space of the system. This channel is called the G-twirling channel and can be treated as a new type of decoherence to the systems \cite{rfrev}. This is an interesting result, due to the fact that the techniques developed in the theory of decoherence free subspaces and subsystems \cite{dfs1,dfs2,dfs3,dfs4,dfs5} can now be used to circumvent this decoherence and succeed in many quantum information tasks in the absence of a SRF, to name but a few, quantum and classical communication \cite{comm,mainref}, quantum key distribution \cite{qkd1,qkd2}, and quantum cryptography \cite{cryp}. 
\\

Multi-partite systems entail two types of degrees of freedom (DoF), namely collective and relative DoFs. The parameters describing the system's relation to some reference external to it are called collective, while those concerned with the relation between the system's parts are known as relative ones. For instance, when one considers a set of two vectors in the Euclidean space, the angle between the vectors is a relative parameter; however, the angle between the bisector of that angle and any of the axes of the Cartesian coordinate system would be a collective DoF. In other words, the relative parameters are the ones that remain invariant under the change of reference frame, while collective parameters change. Similarly, the state of  
quantum  
systems consisting of two qubits is denoted by $\rho_{\gamma,\omega}$, in which $\gamma$ and $\omega$ represent all of its relative and collective DoFs respectively. Based on the above discussion, $\rho_{\gamma,\omega}$ transforms under a global rotation as
\be\label{def}
[D({\bf \Omega})\otimes D({\bf \Omega})]\rho_{\gamma,\omega} [D({\bf \Omega})\otimes D({\bf \Omega})]^\dag = \rho_{\gamma,\omega'}
\ee
where $D({\bf \Omega})\otimes D({\bf \Omega})$ is 
the collective tensor representation of SU(2) on 
the joint Hilbert space $\mathbb{H}\otimes\mathbb{H}$ 
of qubits and ${\bf \Omega}\in$ SU(2) is an arbitrary 
element of the group that rotates both qubits in the same way. 
The global parameters of the state after the transformation change from
$\omega$ to $\omega'$. However, the relative parameters remain unchanged and can therefore 
be used in relational 
encoding for quantum communication tasks. \\

Suppose that a message is encoded into a relative parameter $\gamma$ of a state $\rho_{\gamma, \omega}$. To those who do not have access to a shared Cartesian reference frame with the encoder, the state only entails the relative information; all its collective information is lost and is described by the SU(2)-twirling of $\rho_{\gamma,\omega}$ as
\be
\rho_\gamma=\int [D({\bf \Omega})\otimes D({\bf \Omega})]\rho_{\gamma,\omega} [D({\bf \Omega})\otimes D({\bf \Omega})]^\dag d{\bf \Omega},
\ee
where $d{\bf \Omega}$ is the SU(2) invariant measure. Using a positive operator-valued measure $\{E_\lambda\}$, 
to estimate the relative parameters, 
the prior knowledge of the parameter is updated to a posterior distribution using Bayes's theorem as
\be
p(\gamma|\lambda)=\frac{\mathrm{Tr}(E_\lambda\rho_\gamma)p(\gamma)}{p(\lambda)}.
\ee
The information gain for the 
obtained value of $\lambda$ is given by 
\be
I_\lambda= \int p(\gamma)\log_2 p(\gamma) d\gamma - \int p(\gamma|\lambda)\log_2 p(\gamma|\lambda) d\gamma
\ee 
and the average information gain is given by $I_{avg}=\sum_\lambda P(\lambda)I_\lambda$. We use the average information gain for 
quantifying the performance, i.e., 
the success in estimation of a communicated parameter. 
\\

Bartlett \textit{et al.} \cite{mainref} investigated the problem of communication protocols in the absence of a SRF, when one wants to estimate a relative parameter that is encoded in a pair of product spins. They found that the only relative parameter of a two-qubit product state is the angle between the corresponding vectors of the two qubits in the Bloch sphere and showed that the optimal measurement of the relative parameter can be chosen to be a reference-frame-independent measurement. 
It is natural to investigate the role of entanglement as a ubiquitous property of quantum systems and the relative parameters. More precisely, in the absence of a SRF, shall we encode information in entangled states or in product states? Would one benefit from using entanglement for communication in the absence of the SRF? And at the base of all of these questions, what are the relative parameters of a general two-qubit  state and their physical significance? In this paper we focus on a two-qubit entangled state and answer all of these questions. We show that such a state has three relative parameters and calculate the information gain of the receiver when the value of a bounded continuous parameter is encoded into each of these parameters separately. \\

The structure of the paper is as follows. Section \ref{rel} is devoted to characterization of the relative parameters of a pure entangled two-qubit state. In section \ref{com}, we calculate the average information gain when one of these relative parameters is used for communication. 
We end the paper with a summary and an outlook in section \ref{summ}.
\section{Relative Parameters of a Pure Two-Qubit State}\label{rel}
The only relative parameter of a pure product state of two qubits is 
the angle between the two vectors representing each qubit in the Bloch sphere \cite{mainref}. Here we consider a general pure two-qubit state $|\Psi\ra$, which in the  computational basis is represented as
\be\label{psipsi}
|\Psi\ra = a|00\ra + b|01\ra + c|10\ra + d|11\ra.
\ee
From the normalization of the state, 
$|a|^{2}+|b|^{2}+|c|^{2}+|d|^{2}=1$ and the freedom of a global phase, this state has six real parameters. Invariance under global SU(2) rotations, $D({\bf \Omega})\otimes D({\bf \Omega})$, that have only three parameters, leaves three independent parameters  in the state which are the relative parameters that we are searching for. To find these, one can write 
\be
|\Psi\ra=\sum_{i,j} \Psi_{ij}|i,j\ra,
\ee 
in which $\Psi_{ij}$ can be written as a matrix
\be 
\hat{\Psi}=\left(\begin{array}{cc} a& b \\ c& d\end{array}\right)
\ee
and note that under $D({\bf \Omega})\otimes D({\bf \Omega})$, 
\be
\Psi_{i,j}\lo D({\bf \Omega})_{ik}D({\bf \Omega})_{jl}\Psi_{k,l}
\ee
or in matrix form
\be
\hat{\Psi}\lo \big[D({\bf \Omega})\hat{\Psi} D({\bf \Omega})^T\big].
\ee
The two invariants of this transformation are 
\be\label{invinv}
\det(\hat{\Psi})=ad-bc
\h {\rm and}\h
\mathrm{Tr}(-i\sigma_y\hat{\Psi})=b-c.
\ee
The reason for the  second
equality is the identity $\sigma_yD^T ({\bf \Omega})=D^\dagger ({\bf \Omega}) \sigma_y$ which is manifest if  we write $D({\bf \Omega})= e^{i\Omega_1\sigma_x + i\Omega_2\sigma_y + i\Omega_3\sigma_z}$ and use the commutation relations of the Pauli matrices, $\sigma_y\sigma_i\sigma_y=-\sigma_i^T$.
In fact, a simpler way of deriving these invariants is to directly act on the state (\ref{psipsi}) with a unitary $D({\bf \Omega})=\left(\begin{array}{cc}x&y\\ -y^*& x^*\end{array}\right)$ with $|x|^2+|y|^2=1$ and calculate the new parameters $a',b',c',d'$. It is then easily seen by inspection that the two quantities in Eq. (\ref{invinv}) are invariant. 
 Note that $ad-bc$ is a complex number. Its absolute value is half the concurrence measure of entanglement of the state \cite{concur} which is already known to be invariant under any local operation on qubits. Here we see that its phase is also an invariant of the global rotation $D({\bf \Omega})\otimes D({\bf \Omega})$. Nevertheless, by extracting a global phase from the state $(a,b,c,d)\lo e^{i\eta}(a,b,c,d)$, one can always make $ad-bc$ real and positive and so equal to $|ad-bc|$. So the first invariant $ad-bc$ is nothing but a measure of the entanglement of the two qubits. We are thus left with three real relative parameters in the above two complex quantities as expected.\\

\begin{figure}[h!]
	\centering
	\includegraphics[scale=0.20]{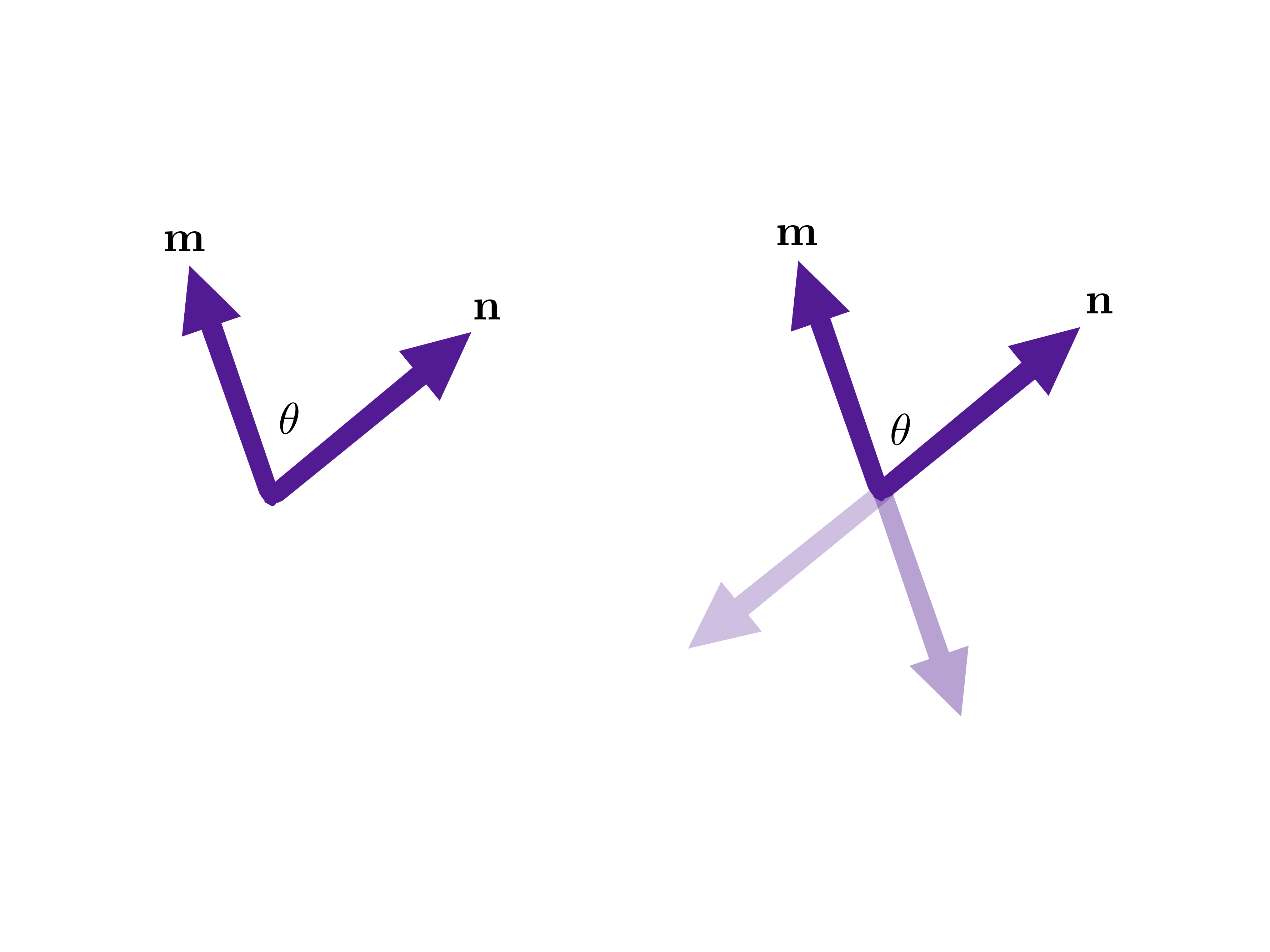}\vspace{-1.3cm}
	\caption{In the absence of a reference frame, a pure product state of two qubits (left) can only carry information in the angle between the two vectors representing the two qubits in the Bloch sphere. An entangled state (right) Eq. (\ref{sch0}), can carry two more parameters. The questions are which parameter is a better carrier and what are the conditions for achieving this optimality.}
    \label{fig:shkl}
\end{figure}

To get a better intuition of the complex invariant $b-c$, we can consider its geometrical expression which is simplified when Alice aligns her coordinate system in a specific way. To see this, let $\rho_A$ be the density matrix of the first qubit with spectral decomposition $\rho_A={\lambda_{m}}^2|{\bf m }\ra\la {\bf m}|$. Expanding the state $|\Psi\ra$ in terms of $|{\bf m}\ra's$ for the first qubit, we find $|\Psi\ra=\sum_m|{\bf m}\ra|\phi_m\ra$, where $\la \phi_m|\phi_l\ra=\delta_{m,l}{\lambda_m}^2.$ This determines $|\phi_m\ra$'s up to a phase. Absorbing the phases into the definition of states, gives the Schmidt decomposition of the state $|\Psi\ra$. 
Note that the Schmidt decomposition does not uniquely determine the state
, e.g. all the Bell states  have the same Schmidt decomposition. 
To fully characterize and distinguish the states,  
we need 
to keep the phases of $|\phi_m\ra$. So the final decomposition of $|\Psi\ra$ would be  
\be\label{sch0}
|\Psi\ra = e^{-i\frac{\psi}{2}}\cos(\alpha)|{\bf m}\ra |{\bf n}\ra +e^{i\frac{\psi}{2}} \sin(\alpha)|{\bf m}^\perp\ra |{\bf n}^\perp\ra, 
\ee
with $ \alpha\in [0,\pi/4]$ and $\psi\in [0,\pi]$. Note that $|\la {\bf m}|{\bf n}\ra|^2=\frac{1}{2}(1+{\bf m}\cdot{\bf n})$, where $\bf n$ and $\bf m$ represent vectors in the Bloch sphere. Based on the representation of the state in Eq. (\ref{sch0}), the invariant parameters under a global rotation $D({\bf \Omega})\otimes D({\bf \Omega})$ are $\alpha, \psi,$ and $\theta$, which is the angle between the two vectors ${\bf m}$ and ${\bf n}$ (see Fig. \ref{fig:shkl}).\\

For simplicity, we assume a coordinate system for Alice such that the two vectors ${\bf m}$ and ${\bf n}$ lie on her \textit{xz} plane. Alice can always choose her coordinate system in this way no matter what the coordinate system of Bob is.  In this coordinate system we get
\begin{eqnarray}\label{n}
|\bf n\ra &=& \cos(\frac{\theta}{2})|0\ra + \sin(\frac{\theta}{2})|1\ra \nonumber \\
|\bf n^\perp\ra &=& - \sin(\frac{\theta}{2})|0\ra + \cos(\frac{\theta}{2})|1\ra.
\end{eqnarray}
Inserting Eqs. (\ref{n}) into Eq. (\ref{sch0}), we obtain the general pure two-qubit state as
\begin{eqnarray}
|\Psi\ra &=& e^{-i\frac{\psi}{2}}\cos(\alpha)|0\ra \left[\cos(\frac{\theta}{2})|0\ra +\sin(\frac{\theta}{2})|1\ra\right] \cr
&+& e^{i\frac{\psi}{2}}\sin(\alpha)|1\ra \left[-\sin(\frac{\theta}{2})|0\ra +\cos(\frac{\theta}{2})|1\ra\right].
\end{eqnarray}
Now, using this form of the two-qubit state, we can see that the relative parameters are
\be
ad - bc = \frac{1}{2}\sin(2\alpha)
\ee
and
\be 
b - c = \sin(\frac{\theta}{2})(e^{-i\frac{\psi}{2}}\cos(\alpha)+ e^{i\frac{\psi}{2}}\sin(\alpha)).
\ee
This indicates that $\alpha$, $\theta$ and $\psi$ can indeed be referred to as the three relative parameters of a pure entangled two-qubit state. \\

\begin{figure}[h!]
	\centering
	\includegraphics[scale=1.1]{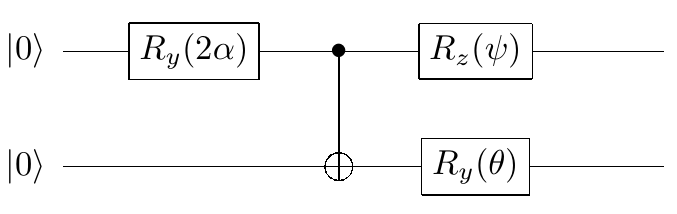}
	\caption{Alice's state preparation circuit. She uses this circuit to independently adjust each of the relative parameters in a suitable way to provide Bob with the highest information gain possible.}
    \label{fig:circuit}
\end{figure}

The fact that the parameters $\alpha, \theta$ and $\psi$ are relative parameters, which can convey information to Bob in the absence of a SRF, means that up to a global rotation $D({\bf \Omega})\otimes D({\bf \Omega})$, Alice should be able to prepare any two-qubit state by application of a quantum circuit based on these three relative parameters. 
The  quantum circuit  in Fig. \ref{fig:circuit} produces the state by which each of the relative parameters can be independently set. \\

Note that the action of this circuit can be written as 
\be\label{circir}
|\Psi\ra=(R_z(\psi)\otimes R_y(\theta))\mathrm{CNOT}(R_y(2\alpha)\otimes I)|0,0\ra.
\ee
If the reference frame of Alice rotates by $D({\bf \Omega})$, this means that the input state of the circuit will change $|0'\ra^{\otimes 2}=D({\bf \Omega})^{\otimes 2} |0,0\ra$ and the gates will change  $R_{n'}(.)=D({\bf \Omega})^{-1}R_n(.)D({\bf \Omega})$.  Inserting this new input state and these new gates into Eq. (\ref{circir}) one finds that the new state is  
  $|\Psi'\ra=D({\bf \Omega})\otimes D({\bf \Omega})|\Psi\ra$ which has the same relative parameters as $|\Psi\ra$, but with new global parameters inherited from $D({\bf \Omega}).$ This means that different reference frames cannot change the relative parameters of the state produced by this circuit.\\
 
We are now faced with the following two questions: \\

i) If Alice wants to communicate the value of a continuous parameter to Bob, to which of the above three relative parameters should she encode this value in order to convey the message with the highest fidelity, i.e., in order for the information gain of Bob to be the highest? \\

ii) Does using the entangled state offer any advantage over a product state? \\
We will answer these questions in the next section. \\
\section{Communication using the relative parameters}\label{com}
In order for Alice to use the relative parameters for communication with Bob, she should encode a message as in Fig. \ref{fig:circuit} and send the two qubits to Bob. Then Bob needs to measure the state and make an estimate of the communicated parameters. We assume that Alice prepares an ensemble of states, all prepared with the relative parameters that encode her message. The optimal measurement done by Bob are the total spin projectors which are $\Pi_0$ and $\Pi_1$ \cite{mainref}. Thus, the probability of projecting on $\Pi_i$, given that $\gamma$ is encoded to the state, is given by   
\be
p(\Pi_i|\gamma)=\mathrm{Tr}(\rho_\gamma \Pi_i).
\ee
Using the Bayesian formalism, Bob can infer the value of $\gamma$ from the following
\be
p(\gamma|\Pi_i)=\frac{\mathrm{Tr}(\Pi_i\rho_\gamma)p(\gamma)}{p(\Pi_i)},
\ee
where $p(\Pi_i)=\int \mathrm{Tr}(\Pi_i\rho_\gamma)p(\gamma)d\gamma$, with $p(\gamma)d\gamma$  a suitable measure over the three component variable $\gamma$. Here $d\gamma$ is a volume element over the manifold of relative parameters the explicit form of which may be difficult to obtain. In all the examples below we are concerned with special forms of the submanifolds where only one of the parameters is changing and the explicit form of $d\gamma$ is simple. Having the posterior distribution over $\gamma$, it is straightforward to calculate the information gain as
\be\label{integrals}
I_{\Pi_i}=\int p(\gamma|\Pi_i)\log_2 \left( \frac{p(\gamma|\Pi_i)}{p(\gamma)}\right) d\gamma
\ee
where the probability $p(\gamma)$ for producing $\gamma$ is known to Alice and Bob. Then we will have
\be
I_{avg}=p(\Pi_0)I_{\Pi_0}+p(\Pi_1)I_{\Pi_1},
\ee
which is the average information gain for Bob given that Alice encodes three real values into $\gamma$. \\

For simplicity, we consider a communication protocol in which Alice uses only one of the relative parameters to encode her message  
and sets the other two parameters to some optimal values that  
maximizes the communicated information. We  assume that prior to the protocol, Alice and Bob agree on the optimal setting, i.e. they both know what values would be used for the relative parameters that are not used for communication. Bob only needs to 
estimate the value of the last remaining relative parameter for finding the message. We also assume that the prior distribution of this parameter is known to both Alice and Bob. \\

For instance, assume that Alice fixes 
 $\theta=\theta_0$ and $\psi=\psi_0$   and encodes her message into the value of the third parameter, $\alpha$. The problem is now to determine the average information gain $I^{(\alpha)}_{avg}(\theta_0,\psi_0)$ and decide what values of $\theta_0$ and $\psi_0$ maximizes this average information gain. In other words, Alice and Bob should determine what fixed values for these two parameters, convey the highest information about the third parameter. We may call this maximum average information gain $I^{(\alpha)}_{avg}(max)$ in this example.
 Similarly, they can use any of $\theta$ and $\psi$ for communication which defines $I^{(\theta)}_{avg}(max),  I^{(\psi)}_{avg}(max)$, 
 respectively. This gives three different approaches for communication using only one relative parameter, and to find the best approach we should determine which of $I^{(\alpha)}_{avg}(max),  I^{(\theta)}_{avg}(max)$, or  $I^{(\psi)}_{avg}(max)$ gives the highest information gain. This will determine which of these three parameters is the best carrier of information in the absence of a SRF. 
 Also to assess the role of entanglement, we  should compare the maximum information gain in each approach with the case where Alice can only send product states and determine whether or not there is any advantage in using entangled states. \\ 

We now note that the state prepared by Alice has three real parameters collectively denoted by $\gamma=(\alpha, \theta, \psi)$. We call this state $|\Psi_\gamma\ra$. When received by Bob, who has no SRF with Alice, it is as if the state has passed through a random global rotation channel and has changed to 
\be
\rho_\gamma = \int [D({\bf \Omega})\otimes D({\bf \Omega})]|\Psi_\gamma\ra \la \Psi_\gamma| [D({\bf \Omega})\otimes D({\bf \Omega})]^\dag d{\bf \Omega}.
\ee
where $d{\bf \Omega}$ is the SU(2) invariant measure. The optimal measurements of Bob are projectors into total spin of the two particles, $\{\Pi_0, \Pi_1\}$. Using the cyclic property of the trace, and then the rotation invariance of total spin projectors, we find
\ba
\mathrm{Tr}(\Pi_i\rho_\gamma)&=&\int \mathrm{Tr} \left\lbrace [D({\bf \Omega})\otimes D({\bf \Omega})]^\dag \Pi_i [D({\bf \Omega})\otimes D({\bf \Omega})]|\Psi_\gamma\ra \la \Psi_\gamma| \right\rbrace d{\bf \Omega}   \cr
&=& \la \Psi_\gamma|\Pi_i|\Psi_\gamma\ra.
\ea
It is worth taking a look at the probabilities of the projections of the state $|\Psi_\gamma\ra$ onto the asymmetric and symmetric subspaces which are
\be \label{proj}
p(\Pi_0|\gamma)=\mathrm{Tr}(\Pi_0\rho_\gamma)=\frac{(1-\cos(\theta))(1+\sin(2\alpha)\cos(\psi))}{4} \\
\ee
and
\be
p(\Pi_1|\gamma)=1-p(\Pi_0|\gamma)
\ee
respectively. One can verify that for product pairs ($\alpha=0$), these probabilities turn into what Bartlett \textit{et al.} used in \cite{mainref}. We use the result in Eq. (\ref{proj}) in each communication scheme to determine the optimal value for one of the two fixed parameters in the following sections. 
We use ``optimal'' for a setting  where, 
the information gain for Bob is maximized, given that Alice prepares a two-qubit state and sends it 
to Bob. \\

We are now ready to investigate the three cases where Alice encodes her message into the value of one of the relative parameters. Since our measurement scheme has only two outcomes the maximum extractable information of it would be 1 bit; this implies that using even all the three parameters for communication, would not increase the maximum information gain. Before presenting the results, we should note what kind of prior probability distribution 
is used by Alice. For each parameter, we consider two different natural distributions, a discrete distribution where the parameter takes two different values with equal probability and a continuous distribution where the parameter is chosen uniformly at random. 
We find the optimal setting for each case. Note that since we are changing the prior probability distribution, we cannot compare the optimal results of the two cases. In other words, although the information 
gain in one scenario may be greater than the other, the optimal setting for each 
case is applicable only to its own circumstances.
In each case, we distinguish the results of these two cases, as shown in Figs. \ref{fig:thetaboth}-\ref{fig:psiboth}-\ref{fig:alphaboth} by the labels "Discrete Distribution" and "Uniform Distribution".
\subsection{Encoding information in $\theta$}
\begin{figure}[h!]
	\centering
	\includegraphics[scale=1]{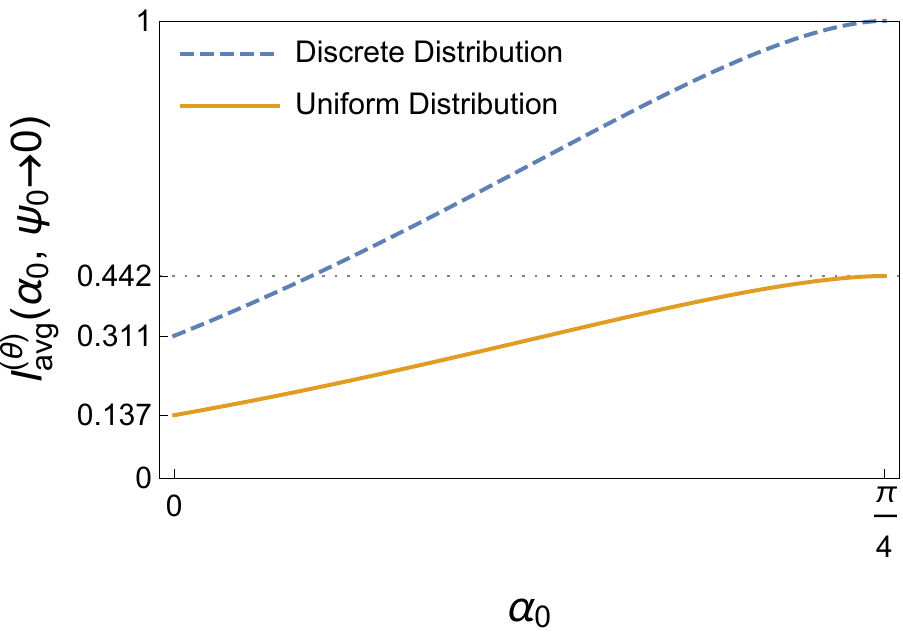}
	\caption{Average information gain (bits) for estimation of $\theta$ in terms of $\alpha_0$ at $\psi_0=0$ for two different prior distributions over $\theta$. The plot shows that the more the entanglement, the higher the average information gain. This also shows that, product states are the worst choice for encoding a message in $\theta$.}
    \label{fig:thetaboth}
\end{figure}

Suppose that Alice encodes her message in  $\theta$. In order to convey the highest amount of information, she should tune the other two parameters into fixed values, say $\alpha_0$ and $\psi_0$. To find the optimal value of $\psi_0$, we can consider the sensitivity of our measurement with respect to the message parameter as 
\be
|\frac{\partial p(\Pi_0|\gamma)}{\partial\theta}|=|\frac{(1+\sin(\theta))(1+\sin(2\alpha_0)\cos(\psi_0))}{4}|.
\ee
It is seen that to increase this sensitivity, Alice should set the parameter $\psi_0=0$ for all $\alpha_0$. Alice does this simply by tuning the gates in her circuit in Fig. \ref{fig:circuit}. To investigate the effect of entanglement and determine $\alpha_0$, we present the results of calculating the average information gain in terms of $\alpha_0$ for $\psi_0=0$. The integrals (\ref{integrals}) are calculated numerically and the results are plotted in Fig. \ref{fig:thetaboth} for the two different prior probability distributions. \\

\textit{\textbf{Uniform Distribution}} Since Alice has fixed the two vectors ${\bf m}$ and ${\bf n}$ in her own frame to lie in the \textit{xz} plane, the prior distribution of $\theta$ is given by $p(\theta)=\frac{1}{\pi}$ for $\theta\in [0,\pi]$. The average information gain is maximized at $\alpha_0=\pi/4$ and equals to 0.442. In the absence of entanglement, $\alpha_0=0$, we find $I^{(\theta)}_{avg} = 0.137$, which shows that entanglement enhances the average information gain almost three times in this protocol. \\

\textit{\textbf{Discrete Distribution}} For the prior $p(\theta=0)=p(\theta=\pi)=1/2$, where the two qubits are either parallel or anti-parallel, the average information gain reaches its maximum at $\alpha_0=\pi/4$, implying that in this case, Bob can exactly retrieve the message which is indeed a classical bit of $1$ for $|\Psi_{\theta=0,\alpha_0=\pi/4,\psi_0=0}\ra=\frac{1}{\sqrt{2}}(|00\ra+|11\ra)$ and zero for $|\Psi_{\theta=\pi,\alpha_0=\pi/4,\psi_0=0}\ra=\frac{1}{\sqrt{2}}(|01\ra-|10\ra)$. In fact, 
Alice and Bob can do classical communication, as previously noted in \cite{comm}. 
Note that this is the maximum average information gain as previously mentioned. Without entanglement ($\alpha_0=0$), the average information gain would be 0.311.
\subsection{Encoding Information in $\psi$}
\begin{figure}[h!]
	\centering
	\includegraphics[scale=1]{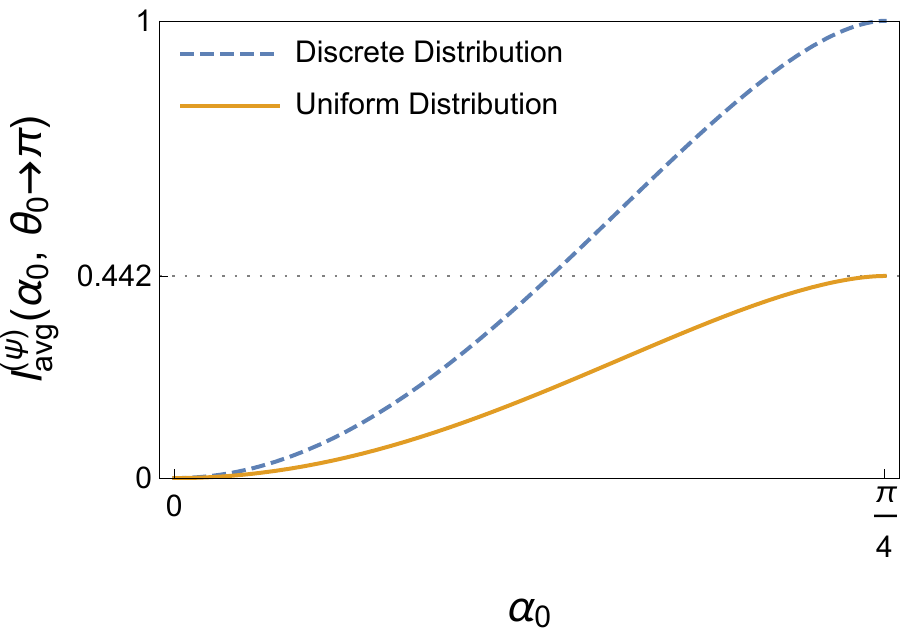}
	\caption{Average information gain (bits) for estimation of $\psi$ in terms of $\alpha_0$ at $\theta_0=\pi$ for two different prior distributions over $\psi$. It is seen that for no entanglement, no information is gained since $\psi$ vanishes for product states. On the other hand, as entanglement increases, so does the average information gain.}
    \label{fig:psiboth}
\end{figure}

When Alice encodes the message in $\psi$, similar to the sensitivity argument of the previous case, we find
\be
|\frac{\partial p(\Pi_0|\gamma)}{\partial\psi}|=|\frac{(1-\cos(\theta_0))(1-\sin(2\alpha_0)\sin(\psi))}{4}|.
\ee
It is inferred that it is best to set $\theta_0=\pi$ for all the values of $\alpha_0$. To find the optimal value of entanglement, the average information gain is considered in Fig. \ref{fig:psiboth} for two different prior probability distributions. \\

\textit{\textbf{Uniform Distribution}} a uniform distribution for $\psi$ is $p(\psi)=\frac{1}{\pi}$ for $\psi\in [0,\pi]$.  The average information gain is maximized at $\alpha_0=\pi/4$, which equals to 0.442. Therefore this relative phase carries the largest information for maximally entangled states. As expected for $\alpha_0=0$, no information is communicated since $\psi$ loses its meaning. \\

\textit{\textbf{Discrete Distribution}} For the discrete distribution $p(\psi=0)=p(\psi=\pi)=1/2$,  the average information gain reaches the value of 1 at $\alpha_0=\frac{\pi}{4}$ which is the maximum extractable information of our measurement. In this case, Bob can exactly retrieve a classical bit encoded in the two states $|\Psi_{\psi=\pi,\alpha_0=\pi/4,\theta_0=\pi}\ra=\frac{1}{\sqrt{2}}(|01\ra+|10\ra)$ and $|\Psi_{\psi=0,\alpha_0=\pi/4,\theta_0=\pi}\ra=\frac{1}{\sqrt{2}}(|01\ra -|10\ra)$. This is because the two states belong to the triplet and singlet subspaces correspondingly
and can be fully discriminated  by Bob's measurement of total spins. Had Alice set $\theta_0$ to any other value, say $\theta_0=0$, for which $|\Psi_{\psi,\alpha_0=\pi/4,\theta_0=0}\ra=\frac{1}{\sqrt{2}}(|00\ra\pm|11\ra)$, this perfect communication could not be achieved.  
\subsection{Encoding Information in $\alpha$}
\begin{figure}[h!]
	\centering
	\includegraphics[scale=1]{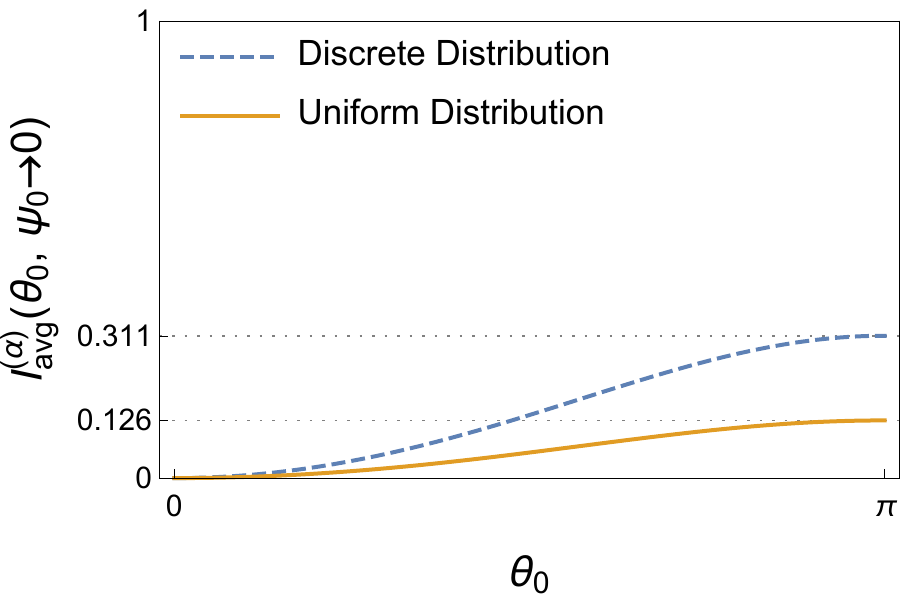}
	\caption{Average information gain (bits) for estimation of $\alpha$ in terms of $\theta_0$ at $\psi_0=0$ for two different prior distributions over $\alpha$. Increasing $\theta_0$ increases the average information gain; therefore, the best states to encode a message in their $\alpha$ parameter are the ones with $\theta_0=\pi$ and $\psi_0=0$.}
\label{fig:alphaboth}
\end{figure}

While encoding the message into $\alpha$ we see that 
\be
|\frac{\partial p(\Pi_0|\gamma)}{\partial\alpha}|=|\frac{(1-\cos(\theta_0))(1+2\cos(2\alpha)\cos(\psi_0))}{4}|.
\ee
Thus, the best setting for this encoding is to set  $\psi_0=0 or \pi$ for all $\theta_0$. In the following, the effect of $\theta_0$ on the average information gain for two different prior probability distributions given that $\psi_0=0$ is investigated. The results are shown in Fig. \ref{fig:alphaboth}. \\

\textit{\textbf{Uniform Distribution}} Taking the prior over $\alpha$ as $p(\alpha)=\frac{4}{\pi}$ for $\alpha\in [0,\pi/4]$, the average information is maximized at $\theta_0=\pi$ and equals to 0.126. In other words, If Alice encodes the value of $\alpha$ into states of the form $|\Psi_{\alpha,\theta_0,\psi_0=0}\ra=\cos\alpha |0,{\bf n}\ra+\sin\alpha |1,{\bf n}^\perp\ra$, in order to provide Bob with the best estimation of $\alpha$, Fig. \ref{fig:alphaboth} implies that $\theta_0$ must be set to $\pi$; this corresponds to the situation in which the value of $\alpha$ is encoded into states of the form $|\Psi_{\alpha,\theta_0=\pi,\psi_0=0}\ra=\cos\alpha |0,1\ra-\sin\alpha |1,0\ra$. \\

\textit{\textbf{Discrete Distribution}} For the prior $p(\alpha=0)=p(\alpha=\pi)=1/2$, Bob's task is to discriminate between maximally entangled states and product states. Looking at Fig. \ref{fig:alphaboth}, we observe that the same trend as the uniform prior is valid with a higher value of the average information gain at the optimal setting for $\theta_0$. \\

Note that for both distributions, no information is transmitted at $\theta_0=0$, that is when Alice encodes the value of $\alpha$ into the state $|\Psi_{\alpha,\theta_0=0,\psi_0=0}\ra=\cos\alpha |0,0\ra+\sin\alpha |1,1\ra$ which regardless of the value of $\alpha$ is always projected onto the triplet subspace. \\
\subsection{Comparing the efficiency of encoding schemes}
It is now interesting to compare the results of the three curves in Figs. \ref{fig:thetaboth}-\ref{fig:psiboth}-\ref{fig:alphaboth}. The maximum amounts of average information gain at the optimal settings are compared in 
table (\ref{fig:tabl}).
\begin{table}[h!]
\centering
\begin{tabular}{|c||*{3}{c|}}\hline
\backslashbox{$I_{avg}(max)$}{Encoding Parameter} & $\theta$ & $\psi$ & $\alpha$ \\\hline\hline
Uniform Distribution & 0.442 & 0.442 & 0.126\\\hline
Discrete Distribution & 1 & 1 & 0.311\\\hline
\end{tabular}
\caption{The highest amount of the average information gain for different encoding schemes.}
\label{fig:tabl}
\end{table}
It is concluded that the the parameter $\alpha$ which determines the amount of entanglement in the state, is the least informative carrier among the three relative parameters. Although $\theta$ and $\psi$ provide the same maximum results, comparing the curves in Figs. \ref{fig:thetaboth}-\ref{fig:psiboth}, we infer that $\theta$ is, in general, a better carrier of information since at low amounts of entanglement, the average information gain for encoding in $\theta$ is higher than for $\psi$. \\

\begin{figure}[h!]
	\centering
	\includegraphics[width=1\columnwidth]{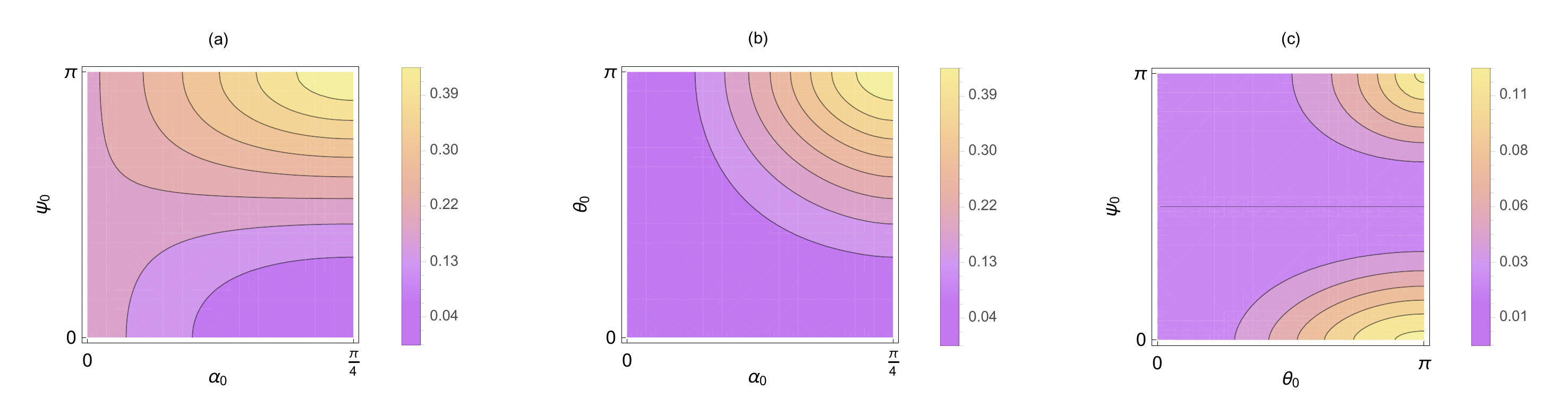}
	\caption{Average information gain (bits) for estimation of $\theta$, $\psi$, and $\alpha$ in terms of the two other parameters in each encoding scenario for the uniform prior distributions over the message parameter. Contours are included in the plots for further clarification. (a): $I^{(\theta)}_{avg}(\alpha_0,\psi_0)$, (b): $I^{(\psi)}_{avg}(\alpha_0,\theta_0)$, (c): $I^{(\alpha)}_{avg}(\theta_0,\psi_0)$.}
\label{fig:uniform}
\end{figure}

\begin{figure}[h!]
	\centering
	\includegraphics[width=1\columnwidth]{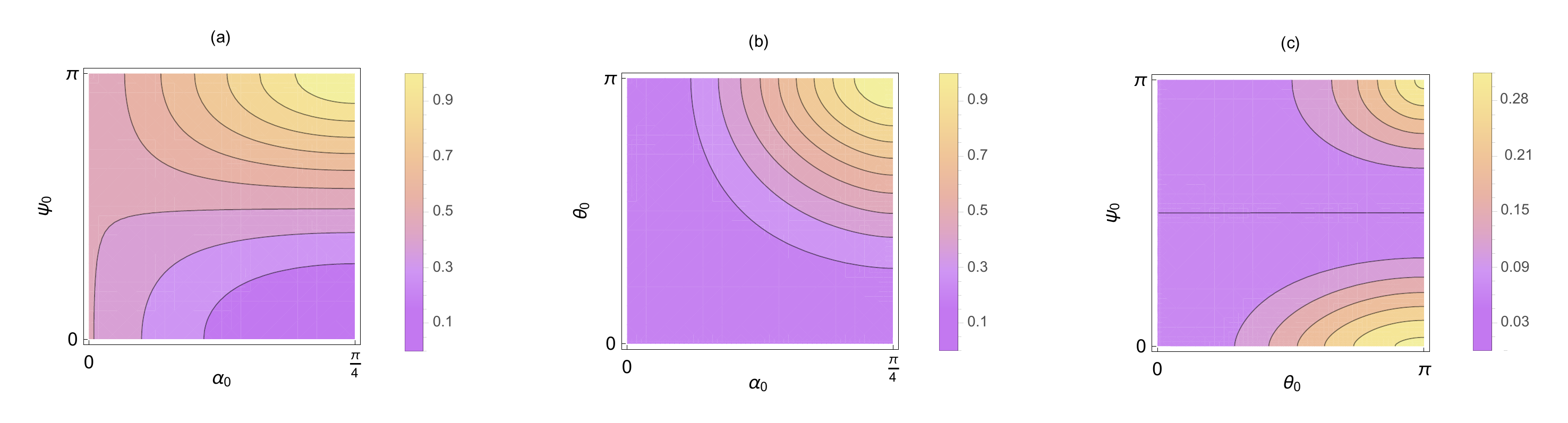}
	\caption{Average information gain (bits) for estimation of $\theta$, $\psi$, and $\alpha$ in terms of the two other parameters in each encoding scenario for the discrete prior distributions over the message parameter. Contours are included in the plots for further clarification. (a): $I^{(\theta)}_{avg}(\alpha_0,\psi_0)$, (b): $I^{(\psi)}_{avg}(\alpha_0,\theta_0)$, (c): $I^{(\alpha)}_{avg}(\theta_0,\psi_0)$.}
\label{fig:discrete}
\end{figure}

It is interesting to observe, in each encoding scenario, how the two other parameters, when they are varied, affect the amount of information communicated. In Figs. \ref{fig:uniform}-\ref{fig:discrete} the average amount of information communicated via the message parameter is plotted with respect to the values of the other two parameters. In fact, Figs. \ref{fig:thetaboth}-\ref{fig:psiboth}-\ref{fig:alphaboth} represent only a slice of these plots. For examples,  Fig. \ref{fig:thetaboth} contains the trends of the average information gain as $\alpha_0$ varies while $\psi_0$ is fixed at $\psi_0=0$ for the two prior distributions; these slices can be tracked inside  Figs. \ref{fig:uniform}(a)-\ref{fig:discrete}(a).

\section{Summary and Outlook}\label{summ}

This paper shows that pure entangled two-qubit states have three relative parameters that are invariant under the change of reference frame, while the pure product two-qubit states entail only one relative parameter. In addition, the result of the communication protocols discussed here, demonstrates that $\theta$, the angle between the two qubits, is the best parameter to convey a message in the absence of a SRF. Our result also shows that the entanglement can be employed to enhance the average information gain of the receiver.  Comparing the highest achievable average information gains for a message encoded into a pure entangled two-qubit state, which are 0.442 and 1 for the uniform and discrete distribution, respectively, with the results of encoding the message into the only relative parameter of a pure product two-qubit state, which are 0.137 and 0.311, respectively, we confirm the positive role of the entanglement in the enhancement of the information communicated.\\ 

It is worthwhile to mention that it is not possible to change the encoding strategy and obtain more information in this protocol. This is due to the fact that the communication protocol in the absence of a SRF is based on the encodings into the relative parameters of a state and the optimal measurement of the relative parameters is shown by Bartlett \textit{et al.} \cite{mainref} to be the total spin projective measurement. For the case of two qubits the Hilbert space of which is made of spin-0 and spin-1 subspaces, any encoding that one uses, would be finally subject to the measurement of its optimal encodings, which are symmetric and asymmetric states. \\

While we have considered only qubits, similar questions can be asked about relative parameters of bipartite qudits and the role that each of their relative parameters plays in communication protocols in the absence of a SRF. 

\section{Acknowledgements} 

This research was partially supported by a grant No. 96011347 from the Iran National Science Foundation and by Grant No. G950222 and G960219 from the research grant system of Sharif University of Technology.

{}

\end{document}